\documentclass[aps,prd,twocolumn,showpacs,preprintnumbers,nofootinbib,amsmath,amssymb,floatfix]{revtex4}

\usepackage{graphicx}
\usepackage{dcolumn}
\usepackage{bm}
\usepackage{amsmath}
\usepackage{comment}

\usepackage[normalem]{ulem}

\newcommand{\ie}{\emph{i.e.} }
\newcommand{\eg}{\emph{e.g.} }

\usepackage{ifthen}
\usepackage{hyperref}
\newboolean{ARXIV}
\setboolean{ARXIV}{true}
\newcommand{\arxiv}[1]{\ifthenelse{\boolean{ARXIV}}{ [\href{http://arxiv.org/abs/#1}{#1}]}{}}

\begin{document}

\title{Phase diagram of the 4D $U(1)$ model at finite temperature.}

\author{Claudio Bonati}
\author{Massimo D'Elia}

\affiliation{Dipartimento di Fisica, Universit\`a di Pisa and INFN, Largo Pontecorvo 3, I-56127 Pisa, Italy}

\date{\today}

\begin{abstract}
We explore the phase diagram of the 4D compact $U(1)$ gauge theory
at finite temperature
as a function of the gauge coupling and of the compactified
Euclidean time dimension $L_t$. 
We show that the strong-to-weak coupling transition, which 
is first order at $T = 0$ ($L_t=\infty$), becomes second order for high temperatures, i.e.
for small values of $L_t$, with a tricritical temporal size $\bar L_t$
located between 5 and 6.
The critical behavior
around the tricritical point explains and reconciles 
previous contradictory evidences found in the literature. 
\end{abstract}

\pacs{
11.15.Ha (Lattice gauge theory),
64.60.Kw (Multicritical points),
}
\maketitle

\section{Introduction}

Discerning the critical behavior taking place at a given
phase transition by numerical simulations can be a very challenging
problem, especially when one has to distinguish between 
a second order transition and a very weak first order transition.

A beautiful example is given by the Abelian compact $U(1)$ gauge theory
in four dimensions in the Wilson discretization:
in this formulation the theory is defined on the 4D Euclidean hypercubic space-time 
lattice, the fundamental dynamical variables are the parallel transports along the
links $U_{\mu}(x)=\exp(ieaA_{\mu}(x))$ ($a$ is the lattice spacing) and the action is:
\begin{equation}\label{action_u1}
S_W= \beta\sum_{x,\mu>\nu}\Big(1-\mathrm{Re}\,\Pi_{\mu\nu}(x)\Big)\ ,
\end{equation}
where the plaquette $\Pi_{\mu\nu}(x)$ is defined as
\begin{equation}\label{plaq0}
\Pi_{\mu\nu}(x)=U_{\mu}(x)U_{\nu}(x+\hat{\mu})U_{\mu}^{\dag}(x+\hat{\nu})U_{\nu}^{\dag}(x) \ .
\end{equation}
It is well known, since the early formulation of the model, that it has two 
different phases: a weak coupling Coulomb phase and a strong coupling 
confined phase. The existence of these two phases was rigorously proved,
for a different form of the action, in Ref.~\cite{Guth}, while for the Wilson action
\eqref{action_u1} evidence of this phase structure is based on numerical simulations
(for recent results see \eg Refs.\cite{DGP, Jersak99, VdP, MKK, P}).
The nature of the transition separating the two 
phases has been debated for a long time, the final conclusion being
that the transition is weak first order~\cite{Jersak83, Arnold01, Arnold03}.
This conclusion is important since it implies that no continuum Quantum Field Theory 
can be defined for the $U(1)$ gauge theory at the critical point, since the correlation 
length stays finite.

More recently, a finite temperature version of the model has been considered,
in which the Euclidean temporal size $L_t$ is compactified and kept fixed,
while the thermodynamical limit is reached by sending to infinity 
the spatial sizes, $L_s \to \infty$. The model shows a non-trivial phase 
structure in the $1/L_t - \beta$ plane. For $L_t = 1$, the system gets 
exactly decoupled into a 3D $XY$ model and a 3D $U(1)$ lattice 
gauge theory: the former undergoes a second order phase transition
at $\beta_c = 0.4541652(5)(6)$ (see Ref.~\cite{Campostrini06}) while the latter is known to be always in the 
confined phase (see Refs.~\cite{Polyakov76, Gopfert81}), therefore a single phase 
transition is expected, in the 3D XY universality class. Such phase transition extends
also to $L_t \neq 1$ and corresponds to the spontaneous breaking
of a symmetry analogous to that of the $XY$ model, \ie a global 
$U(1)$ symmetry. This is generated by gauge transformations which are 
periodic in time only up to a given global phase, and an order parameter for 
this symmetry is the Wilson line taken along
the Euclidean temporal direction, \ie the Polyakov loop. For
$L_t \to \infty$, the breaking of this symmetry corresponds to the 
confining-to-Coulomb transition of the 4D theory, which is known to be first order.
A possible further transition line may be present, in principle, separating
a weak-coupling phase with spatial confinement at small $L_t$ from 
the zero-temperature Coulomb phase, however evidence has been presented
in Ref.~\cite{vettorazzo} that such line could be actually placed at
$L_t \to \infty$.

A natural question is how the order of the $U(1)$-breaking transition 
changes as one moves from $L_t = 1$, where it is second order, to 
$L_t = \infty$, where it is first order. Notice that, for every value of $L_t$, 
the fact that an exact global symmetry exists guarantees that
a true phase transition must be found, associated
with its spontaneous breaking.

This issue has been discussed for the first time in Ref.~\cite{vettorazzo},
where it has been proposed that the transition may stay first order
at least for $L_t$ down to $L_t = 6$, turning into second order for lower
$L_t$. The evidence in that case was based on the analysis 
of the time histories on large lattices (up to $6\times 60^3$), showing 
clear signs of metastability. However, due to computational limitations,
statistics presented in Ref.~\cite{vettorazzo} were not enough 
to reach a definite conclusion.

On the contrary, 
in a subsequent investigation, the authors of Ref.~\cite{berg} have suggested
that a finite size scaling analysis gives evidence for a second order
transition at $L_t = 6$, with critical indexes similar to that of a Gaussian
point (see also Ref.~\cite{berg_lat}). The authors of Ref.~\cite{berg} find
Gaussian indexes, instead of the expected XY indexes, 
also for smaller values of $L_t$, down to $L_t = 4$. 
The problem in the analysis of Ref.~\cite{berg}, as pointed
out by the authors themselves and as we will clarify later, 
is the limited number of available spatial sizes, going up to 
$L_s = 18$, which does not permit a correct extrapolation to the 
thermodynamical limit.

In the present study we solve the issue, reconciling evidence
from Ref.~\cite{vettorazzo} and \cite{berg}. In particular,
we will show that the transition is indeed first order down
to $L_t = 6$, turning into second order for lower values 
of $L_t$. The separation between the two different regimes
is marked, as it always happens in these cases, by a tricritical point,
which, even if not exactly located at an integer value of $L_t$,
influences the scaling of nearby values of $L_t$. As a consequence, 
the correct scaling on such points, first order or second order, is not visible
until large enough values of the spatial size $L_s$ are reached.
On small lattices, instead, the true scaling behavior is obscured by a fake tricritical 
scaling, which has indeed the mean field tricritical (Gaussian) indexes observed in Ref.~\cite{berg}.
This variation of the scaling behavior with the lattice size is typical of tricritical points and is
observed in Monte Carlo simulations of many other models, going from QCD at finite baryon chemical potential 
(see e.g Refs.~\cite{DES, dFP, BCDES, BdFDEPS}) to the Potts model in an external magnetic field (Ref.~\cite{BDE}). 
The simplest physical systems which exhibit tricritical behavior are probably fluid mixtures but, as should 
also be clear from the previous discussion, tricritical phenomena appear to be ubiquitous in physics (see \eg
Ref.~\cite{LawSarb}). In relation to our study, particularly intriguing is the presence of tricritical points
in superconductors (see Refs.~\cite{Kleinert82, Kleinert06}): Ginzburg-Landau theory of superconductivity 
is based on an $U(1)$ gauge theory (with matter) and the tricritical behavior is triggered by vortex line defects, 
which are present also in the lattice $U(1)$ gauge theory.

The paper is organized as follows. In Section~\ref{setup} we present
a general overview of the quantities and the methods adopted
to probe the critical behavior around the transition;
in Section~\ref{results} we present and discuss numerical results
obtained for various values of $L_t$; 
finally, in Section~\ref{concl}, we draw our conclusions.

\section{Observables and numerical analysis setup}\label{setup}

\begin{table}[bt!]
\begin{tabular}{|c|c|c|c|c|c|}
\hline & $\nu$ & $\gamma$ & $\alpha$ & $\gamma/\nu$ & $\alpha/\nu$\\
\hline $3D$ XY & 0.6717(1) & 1.3178(2) & -0.0151(3) & 1.962 & $-$0.0225\\
\hline Tricritical & 1/2 & 1 & 1/2 & 2 & 1\\
\hline $1^{st}$ Order & 1/3 & 1 & 1 & 3 & 3\\
\hline
\end{tabular}
\caption{Critical exponents (see \cite{Campostrini06} and \eg \cite{landau, pelissvic}).}\label{CRITEXP}
\end{table}

In lattice gauge theories, the natural observables are mean values of path ordered 
products of link variables. The simplest path 
that can be used is the boundary of an elementary square, the corresponding 
observable being the mean value of the plaquette operator
\begin{equation}\label{plaq}
W=\frac{1}{6L_tL_s^3}\sum_x \mathrm{Re}\,\Pi_{\mu\nu}(x)\ ,
\end{equation}
where $\Pi_{\mu\nu}(x)$ is defined in Eq.~\eqref{plaq0}. $W$ is proportional (up to an additive
constant) to the energy density of the system.

The other variable that will be used is the Polyakov loop: in this case the path is a
straight line along the temporal direction, which closes because of the periodic 
boundary conditions in the temporal direction, 
\begin{equation}\label{poly}
P=\frac{1}{L_s^3}\sum_{\mathbf{x}} \prod_{t=0}^{L_t-1} U_0(t, \mathbf{x})\ ,
\end{equation}
where $\mathbf{x}$ denotes a generic point of the $t=0$ slice of the lattice. As noted in the introduction
this observable is an order parameter for the breaking of a global $U(1)$ symmetry. For 
larger gauge groups a non-vanishing $P$ value signals the breaking of the center symmetry, 
which in our case is just the group itself, being $U(1)$ abelian. 
The fact that the transition of the ($3+1$)D $U(1)$
gauge theory, if second order, is in the universality class of the 3D $XY$ model, can be viewed as a particular realization of the Svetitsky-Yaffe conjecture (Ref.~\cite{svya}).

Our goal is to show that the order of the deconfinement transition changes by increasing $L_t$. 
To this aim we need to study, at fixed $L_t$, the critical behavior of the system for 
$L_s\to\infty$, in order to discriminate between a first and a second order transition.
We will thus study the susceptibilities of $W$ and $P$, defined by
\begin{equation}
C_V = L_t L_s^3\ (\langle W^2 \rangle - \langle W \rangle^2) \label{susc_plaq}
\end{equation}
and 
\begin{equation}
\chi = L_t L_s^3\ (\langle |P|^2 \rangle - \langle |P| \rangle^2) \, . \label{susc_poly}
\end{equation}

Near the phase transition, the scaling of these two quantities as functions of the spatial 
size $L_s$ is given (up to additive analytic contributions) by
\begin{equation}
\begin{aligned}
C_V &\sim L_s^{\alpha/\nu}\ f_1 (t L_s^{1/\nu})  \\
\chi &\sim L_s^{\gamma/\nu}\ f_2 (t L_s^{1/\nu})\ ,
\end{aligned}
\end{equation}
where $t \equiv (T - T_c)/T_c$ is the reduced temperature and the $f_i$'s are universal scaling functions,
\ie they depend only on the universality class of the (second order) transition. From these relations it follows 
in particular that the scaling of the height of the peaks is governed by the exponents $\alpha/\nu$ and 
$\gamma/\nu$ respectively.
The critical indexes which will be relevant in the following are those of the 3D $XY$ model, 
the tricritical and the first order ones. Their numerical values are reported
for convenience in Table~\ref{CRITEXP}. 

\begin{figure}[h]
\includegraphics*[width=0.46\textwidth]{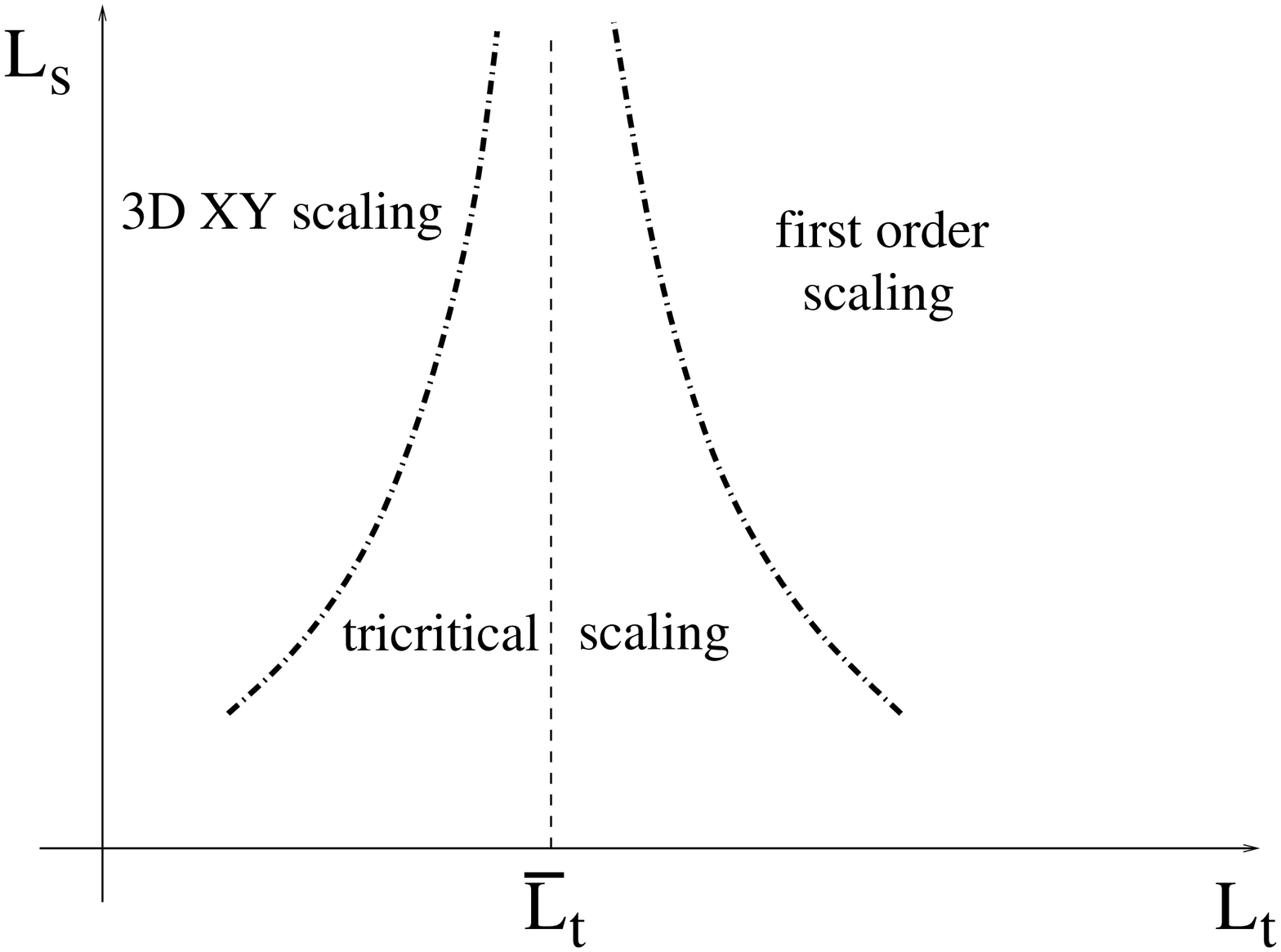}
\caption{Plot of the regions where the different critical behaviors are expected to be seen 
on finite volumes. The sector where tricritical scaling dominates
shrinks to zero in the thermodynamic limit but can be very large for small volumes.}
\label{triscal}
\end{figure}

It was noted in Ref.~\cite{BDE} that it is increasingly difficult to correctly identify 
the order of the transition as we approach a tricritical point, since larger and larger volumes are needed 
in order to disentangle the true thermodynamic behavior from a fictitious tricritical
scaling for small volumes (see Fig.~\ref{triscal} for a pictorial representation).
In fact it can be shown \cite{BDE, pelissvic, BinderDeutsch} that the spatial lattice
size required for an unambiguous identification of the transition scales as
\begin{equation}\label{triccrossover}
L_s\sim A |L_t-\bar{L}_t|^{-1}
\end{equation}
where $A$ is a constant and $\bar{L}_t$ is the tricritical point value.

It thus follows that, in order to locate with precision the tricritical point, we 
cannot simply look for regions of the parameter space where a tricritical scaling is
observed.  A better strategy is to study observables which quantify the strength 
of the first order transition, like the discontinuities at the transition, and analyze 
their variation as a function of the external parameter ($L_t$ in our case)
in order to extrapolate the point where they vanish, which is just the tricritical point. 

We will denote by $\Delta_{W}$ and $\Delta_P$, respectively, the discontinuities across the (possible) first 
order transition of the mean plaquette and of the Polyakov loop. These quantities can be 
estimated by looking at the scaling of the maxima of the relative susceptibilities at the transition:
if the volume is large enough (\ie if $L_s$ is much larger than the correlation length) we have
\begin{equation}
\begin{aligned}\label{suscmax}
& (C_V)_{max}\sim L_tL_s^3\,\Delta_{W}^2 \\
& (\chi)_{max}\sim L_tL_s^3\,\Delta_{P}^2 \ .
\end{aligned}
\end{equation}
Another estimator for $\Delta_{W}$ is the Binder-Challa-Landau cumulant~\cite{Challa},
which is defined by
\begin{equation}
B_4=1-\frac{\langle W^4\rangle}{3\langle W^2\rangle ^2}\ .
\end{equation}
It can indeed be shown (see \eg Ref.~\cite{LeeKosterlitz}) that near a transition $B_4$ develops a minimum, 
whose depth scales like
\begin{eqnarray}
B_4|_{min} \simeq \frac{2}{3}-\frac{1}{3}\left(\frac{\Delta_{W}}{\epsilon}\right)^2 \label{Bmin}
\end{eqnarray}
where $\epsilon=\frac{1}{2}(W^++W^-)$ and $W^{\pm}=\lim_{\beta\to\beta_c^{\pm}}\langle W\rangle$. 
In particular, the thermodynamical limit of  $B|_{min}$ is 
strictly less than \(2/3\) if and only if a discontinuity is present. 
In order to simplify the notation in the following we will adopt the shorthand $B=\frac{2}{3}-B_4|_{min}$.

The discontinuities $\Delta_{W}$ and $\Delta_P$ decrease as we approach the tricritical temporal size
$\bar{L}_t$ from the first order side (\ie from $L_t>\bar{L}_t$) and the leading order behavior is
(see Ref.~\cite{LawSarb} or \cite{Sheehy} for a brief summary of the main results)
\begin{equation}
\Delta_{W} \propto \sqrt{L_t - \bar{L}_t} \label{deltae_beh}
\end{equation}
and
\begin{equation}
\Delta_P \propto \sqrt{|(L_t - \bar{L}_t)\log(L_t-\bar{L}_t)|}\ . \label{delta_beh}
\end{equation}
In systems for which the tricritical behavior is triggered by a continuous variable (like \eg the 
one studied in Ref.~\cite{BDE}) it is typically possible to approach the tricritical point close enough to 
observe a scaling of the form in Eqs.~\eqref{deltae_beh}-\eqref{delta_beh}. In the case at study, the 
relevant variable is discrete and it is not possible to observe such a scaling behavior. A different strategy could 
be to study the 4D $U(1)$ gauge theory with an asymmetric coupling in the temporal direction, 
which effectively reduces the temporal size in a continuous way. However,
in the present investigation we will limit ourselves to the isotropic
case already studied in Refs.~\cite{vettorazzo} and \cite{berg}.

\begin{figure}[h]
\includegraphics*[width=0.46\textwidth]{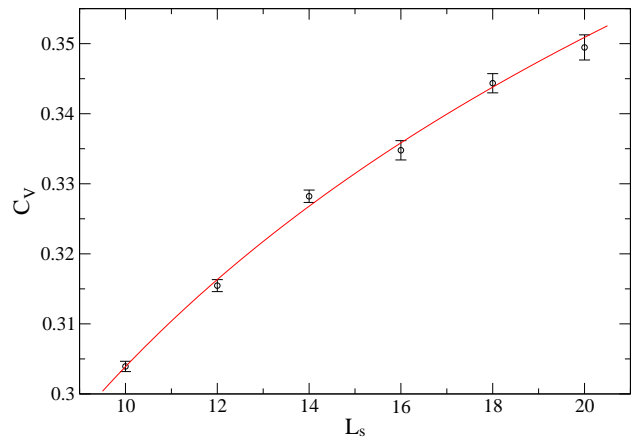}
\caption{Values of the $C_V$ maxima for $L_t=2$. The line is a fit with a function of the 
form $a+b L_s^{\alpha/\nu}$, where for $\alpha$ and $\nu$ the 3D $XY$ values are used 
($\chi^2/\mathrm{d.o.f.}\approx 1.2$).}
\label{chi_plaq_t2}
\end{figure}

\begin{figure}[h]
\includegraphics*[width=0.46\textwidth]{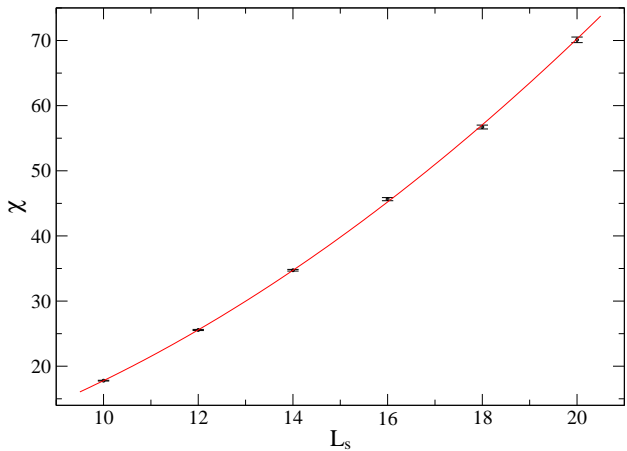}
\caption{Values of the $\chi$ maxima for $L_t=2$. The line is a fit with a function of the 
form $a+b L_s^{\gamma/\nu}$, where for $\gamma$ and $\nu$ the 3D $XY$ values are used 
($\chi^2/\mathrm{d.o.f.}\approx 1.1$).}
\label{chi_poly_t2}
\end{figure}

\section{Numerical Results}\label{results}

Since we are going to study the region of first order transitions near the 
tricritical point, the strength of the discontinuities will never be large enough to 
justify the use of algorithms specifically designed for strong first orders,
like \eg the multicanonical algorithm \cite{BergNeuhaus}. For this reason, we adopt
a mixture of heatbath \cite{Moriarty} and overrelaxation \cite{Creutz} updates in the ratio
$1:5$. Data have been analyzed by means of standard jackknife and multi-histogram reweighing 
algorithms (see \eg Refs.~\cite{BergMCMC, NewmanBarkema} and \cite{FS1, FS2}).
In all the cases $\mathcal{O}(10)$ $\beta$ values were simulated, and for each one $\mathcal{O}(10^3)$
independent measures have been performed (which amounts to $\mathcal{O}(10^6\div 10^7)$ elementary 
update sweeps).

Our first aim is to show that for small $L_t$ the transition is second order in the 
3D $XY$ universality class. In order to prove that, we studied the system with the smallest 
nontrivial value of $L_t$, namely $L_t=2$. 
In Figs.~\ref{chi_plaq_t2} and \ref{chi_poly_t2} the maxima of $C_V$ and $\chi$,
respectively, are plotted for increasing $L_s$ values, and a nice agreement with the theoretical
expectations based on the 3D $XY$ exponents is observed.

However, the right question to ask is to what
extent we can distinguish the critical 
behavior dictated by the 3D $XY$ indexes from that
corresponding to tricritical indexes (coinciding with Gaussian
indexes), so as to exclude the latter. We do not learn
much from the scaling of $\chi$ in Fig.~\ref{chi_poly_t2}:
if we leave the critical exponent as a free parameter we get
$\gamma/\nu=1.939(45)$, which is compatible with both behaviors
(see Table~\ref{CRITEXP}).

Instead, if we look at the plaquette susceptibility, Fig.~\ref{chi_plaq_t2},
we learn that 
on small lattices (\ie $L_s\lesssim 14$) the behavior is still
compatible with Gaussian indexes ($\alpha/\nu = 1$), while on larger
lattices deviations are significant, indicating the need for  
$\alpha/\nu<1$; the 3D $XY$ exponent, on the contrary, describes well
all the explored range of $L_s$. 

The outcome for $L_t = 2$ is therefore that on small lattices
the transition  can be (erroneously) associated with Gaussian critical exponents, while on large enough lattices 
the transition is clearly described by the 3D $XY$ indices, 
with no significant contaminations from tricritical scaling.

\begin{figure}[h]
\includegraphics*[width=0.46\textwidth]{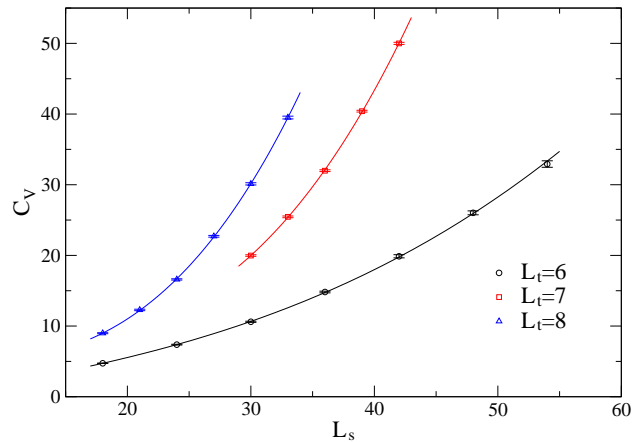}
\caption{Values of the $C_V$ maxima for $L_t=6, 7, 8$. The lines are fits with a function of the 
form \eqref{fit_funct}, with $\lambda_t=1$ 
($\chi^2/\mathrm{d.o.f.}\approx 0.5, 1.1, 0.7$ for $L_t=6,7,8$ respectively).}
\label{chi_plaq_first}
\end{figure}

\begin{figure}[h]
\includegraphics*[width=0.46\textwidth]{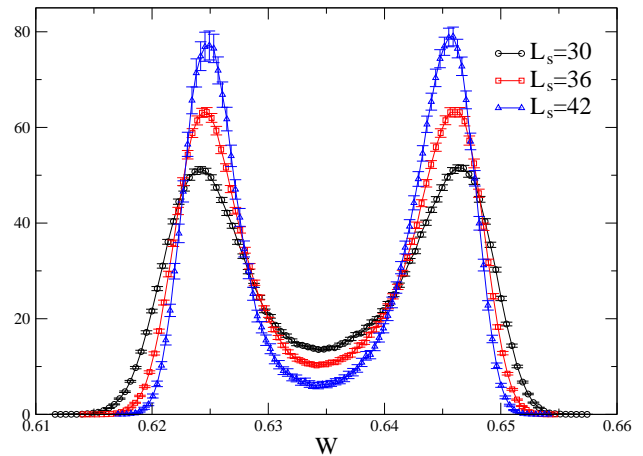}
\caption{Histograms of the plaquette distribution at transition for $L_t=7$ and three different spatial extent.}
\label{histo}
\end{figure}

\begin{figure}[h]
\includegraphics*[width=0.46\textwidth]{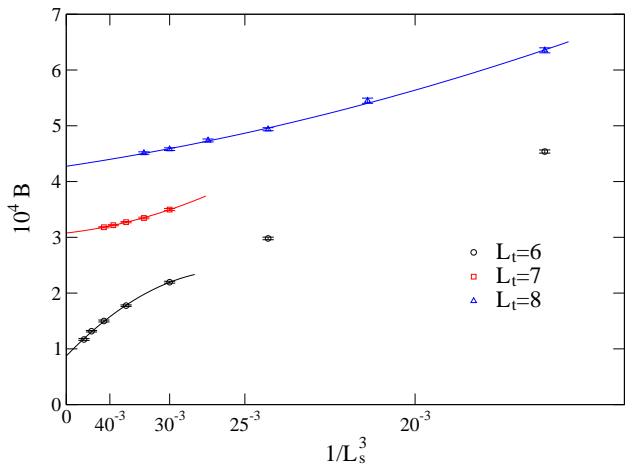}
\caption{Values of $B=2/3-B_4|_{min}$ for $L_t=6, 7, 8$. The lines are fits with $a+bx+cx^2$
($\chi^2/\mathrm{d.o.f.}\approx 1.0, 0.2, 0.7$ for $L_t=6,7,8$ respectively).}
\label{binder_plaq_first}
\end{figure}

The situation is different for $L_t = 6$ or larger, where 
the transition turns out to be first order,
however we had to study lattices up to $L_s=54$ 
to clearly determine the true critical behavior.
Even on the larger lattices, indeed,
the tricritical contribution, although no more dominant, is 
nevertheless significant. To correctly describe the maxima of the susceptibilities we had to 
use a function of the form
\begin{equation}\label{fit_funct}
f_{\chi}(L_s)=a+bL_tL_s^{\lambda_t}+cL_tL_s^{3} 
\end{equation}
where $L_s^3$ is the behavior expected at a first order transition, while the sub-leading term $L_s^{\lambda_t}$ is a
correction with the exponent of the tricritical case ($\lambda_t=1$ and $2$ for the plaquette
and Polyakov loop respectively). 
The $L_t$ multiplicative terms follow from the normalization of the susceptibilities in 
Eqs.~\eqref{susc_plaq}-\eqref{susc_poly}.
Even for the plaquette susceptibility, for which $\lambda_t=1$, the sub-leading term 
is in fact the larger one for $L_s$ up to $\approx 45$, and contributes $\approx 88\%$ of the total singular part 
for $L_s=18$ (which was the largest spatial size explored
in Ref.~\cite{berg}).

The agreement between the data and the fits of the form \eqref{fit_funct} is good, as shown in 
Fig.~\ref{chi_plaq_first} for the susceptibility of the plaquette and for $L_t=6,7,8$. 
From Fig.~\ref{chi_plaq_first} it can be noticed that the first order transition gets stronger with increasing $L_t$, 
as theoretically expected. 
In Fig.~\ref{histo} we show the nice double peak structure which develops in the plaquette distribution at the
transition temperature, which gets more and more pronounced as the thermodynamical limit is approached.

A consistent picture emerges from the study of the Binder cumulant: in Fig.~\ref{binder_plaq_first} the 
values of $B=2/3-B_4|_{min}$ are shown for $L_t=6,7,8$ together with parabolic fits. The reason for the parabolic fits is
that finite size corrections to $B$ are known to be analytic in $1/V$ for first order transitions, see 
Ref.~\cite{LeeKosterlitz}. Remembering Eq.~\eqref{Bmin}, it is evident from Fig.~\ref{binder_plaq_first} 
that the first order transition 
gets stronger as $L_t$ increases.

Following the strategy outlined in Section~\ref{setup}, we will now determine the parameters which fix
the strength of the first order transition, in order to extrapolate the critical value $\bar L_t$ at which the 
first order disappears. The parameters are the latent heat, or equivalently the 
minimum of the Challa-Landau-Binder cumulant defined in Eq.~(\ref{Bmin}), and the 
gap of the order parameter.

\begin{table}[t]
\begin{tabular}{|l|l|l|l|}
\hline 
$L_t$ &  $\Delta_{W}^2$          &  $\Delta_P^2$            & $B$ \\ \hline
$6$   &  $2.04(5)\times 10^{-5}$  &  $1.5(3)\times 10^{-3}$  & $8.7(3)\times 10^{-5}$  \\ \hline 
$7$   &  $1.06(6)\times 10^{-4}$  &  $4.98(8)\times 10^{-3}$ & $3.08(2)\times 10^{-4}$ \\ \hline
$8$   &  $1.39(3)\times 10^{-4}$  &  $3.64(7)\times 10^{-3}$ & $4.27(4)\times 10^{-4}$ \\ \hline
\end{tabular}
\caption{Estimated values for the parameters $\Delta_{W}$, $\Delta_P$ (defined by Eq.~\eqref{suscmax}),
and $B=2/3-B_4|_{min}$ (see Eq.~\eqref{Bmin}).}\label{gaps}
\end{table}

The latent heat and the gap of the order parameter can be extracted from the large volume limit of 
the maxima of the susceptibilities, see Eq.~(\ref{suscmax}). Since we have seen that the tricritical component
significantly contributes to the susceptibilities also for large volumes, 
we decided to adopt the coefficient $c$ in Eq.~\eqref{fit_funct}
as an estimator of the gap,
in order to clearly disentangle the 
first order contribution from the tricritical one. 

\begin{figure}[t]
\includegraphics*[width=0.46\textwidth]{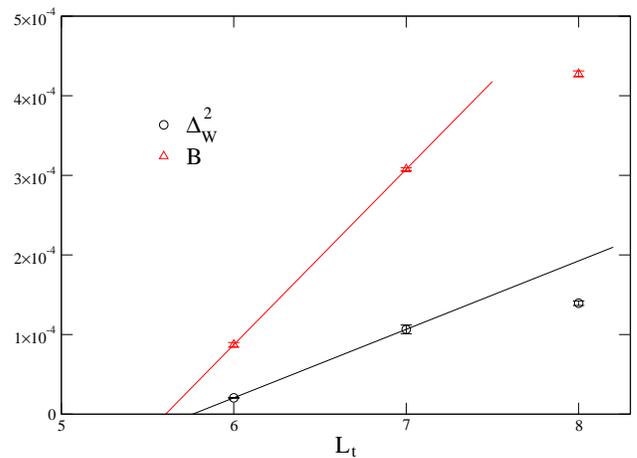}
\caption{Estimates of the endpoint: the lines are linear interpolations of the $L_t=6,7$ data (see text).}
\label{endpoint}
\end{figure}

The values of $\Delta_{W}^2$ and $\Delta_P^2$ estimated from the fits are reported in Table~\ref{gaps}. 
The behavior of $\Delta_{W}$ is qualitatively consistent with the theoretical expectations on the first order
getting stronger with increasing $L_t$, while $\Delta_P$ is non-monotonic with respect to $L_t$. This can be 
explained by noting that, while the plaquette is a local observable, the Polyakov loop is an extended object 
directly related to $L_t$. As a consequence, the gap in the Polyakov loop at the transition cannot in general be directly 
connected to the strength of the transition; on the contrary, we expect that
the gap vanishes in the limit $L_t\to\infty$, since in such limit
the Polyakov loop tends to zero also in the deconfined phase.
It is nevertheless surprising that the results obtained for just three $L_t$ values 
(and in fact for the three smallest values for which a first order transition is present, see later) 
are sufficient to expose this non-monotonic behavior.

By fitting the Binder cumulant values we can estimate the value of $B$ in the thermodynamical limit
(see again Table~\ref{gaps} for the results). This is just $\Delta_{W}^2$ multiplied by a function of $L_t$ which is 
expected not to vanish nor to have singularities at the tricritical point (see Eq.~\eqref{Bmin}).
The behavior of $B$ near the tricritical point should thus be the same as that of $\Delta_{W}^2$.

The expected scaling of $\Delta_{W}$ in the neighbourhood of the tricritical point is given by 
Eq.~\eqref{deltae_beh}, \ie $\Delta_{W}^2$ (and thus also $B$) should depend linearly on $L_t$.
However, as previously noted, we cannot expect to be able to really observe this scaling, since $L_t$ assumes only 
discrete values and thus we are not allowed to approach the tricritical point with arbitrary precision. 

As can be seen in Fig.~\ref{endpoint}, strong deviations from the leading linear behavior are indeed clearly visible
even on the data for three consecutive $L_t$ values. 
However the discrete nature of $L_t$, which introduces such complication,
also gives us the possibility to avoid 
a precise determination of the tricritical value $\bar{L}_t$:
we only need to determine the two consecutive integers such that 
$\bar{L}_t$ is located in between, such task may be unfeasible
only in the unlucky situation in which $\bar{L}_t$ itself is very close 
to an integer.

$B$ and $\Delta_{W}^2$ in Fig.~\ref{endpoint} appear to be concave functions of $L_t$, hence we can obtain an 
underestimate for $\bar{L}_t$ 
by imposing the vanishing of the linear interpolation obtained 
from the two lowest values of $L_t$, \ie $L_t = 6$ and 7. 
As can be seen from Fig.~\ref{deltae_beh},
this underestimate is significantly larger than $L_t=5$. On the other hand,
we know that for $L_t=6$ the transition is first order.
Therefore, we can safely conclude that $5 < \bar L_t < 6$.

A direct check of this statement could be obtained by studying the $L_t=5$ 
system and verifying the presence of  a second
order transition in the 3D $XY$ universality class. 
Such a direct check seems however to be 
more difficult than the corresponding 
one performed on the first order side for $L_t=6$.
Indeed, the critical exponents
of the 3D $XY$ class are smaller than the tricritical ones 
(the opposite happens instead for first order exponents),
so that a direct identification of the universality class 
to which the $L_t = 5$ transition belongs,
might be much harder.

\section{Conclusions}\label{concl}

In the present study we have investigated the 
4D compact $U(1)$ gauge theory at finite temperature,
i.e. on lattices with a finite Euclidean temporal dimension $L_t$,
compactified with periodic boundary conditions.
In particular, we have determined the order of the
strong-to-weak coupling transition as a function of $L_t$.

The transition is first order for $L_t \geq 6$ and second order,
in the universality class of the 3D $XY$ model, for 
$L_t \leq 5$. A tricritical point is thus present 
at a non-integer value  $\bar L_t$, with $5 < \bar L_t < 6$.
The tricritical scaling associated with this point influences,
for nearby values of $L_t$, the scaling of relevant 
susceptibilities in a range intermediate spatial sizes $L_s$,
thus explaining and reconciling 
the contradictory evidence reported in previous
literature~\cite{vettorazzo, berg}.

Evidence for the presence of a first order transition
has been direct on all explored lattices, $L_t = 6,7,8$.
Evidence for a second order transition has been direct 
for $L_t = 2$ and indirect for $L_t = 3,4,5$, in particular based 
on the vanishing of the first
order gap for lattices with $L_t \leq 5$.  

An accurate verification of the correct scaling 
around the tricritical point has not been possible,
due to the discrete nature of the temporal extension $L_t$.
That could be done in a different setup, namely 
working on anisotropic lattices, with two different couplings
for spatial and temporal plaquettes, and approaching 
the tricritical point by tuning the temporal 
gauge coupling. We leave that to future studies.

Finally, let us remark, following the conclusions 
of Ref.~\cite{vettorazzo}, that speaking of  finite 
temperature, in the case of the compactified 
4D $U(1)$ gauge theory that we have studied, is not
completely appropriate, in particular in connection with
the continuum limit of the theory. Indeed, the $a \to 0$ 
limit is possible only in correspondence of second 
order transition points, but since these are limited to 
values of $L_t \leq 5$, a true continuum limit at fixed
physical temperature $T = 1/(L_t a)$ is not possible.
This is at variance with ordinary non-Abelian gauge theories
at finite temperature, where instead 
one can send at same time $L_t \to \infty$
and $a \to 0$, keeping a fixed physical value of $T$.

\section*{Acknowledgments}

Numerical simulations have been performed on GRID resources provided by INFN
and in particular on the CSN4 cluster.
We thank A.~Bazavov, Ph. de Forcrand, A. Papa for useful discussions.
We thank the Galileo Galilei Institute for Theoretical Physics
for the hospitality offered during the workshop ''New Frontiers in
Lattice Gauge Theories".

\appendix

\section{Numerical data}

In this appendix we report the numerical data used in the analysis. The pseudo-critical 
coupling $\beta_{pc}$ is defined by the position of the peak in the Polyakov loop susceptibility.

\begin{table}[h!]
\begin{tabular}{|l|l|l|l|l|}
\hline 
$L_s$ &  max $C_V$ & min $B_4$ & max $\chi$ & $\beta_{pc}$\\ \hline
10 &  0.30392(73) &  0.6658711(18)  &  17.786(53) &  0.8956(12)    \\ \hline
12 &  0.31546(85) &  0.6661848(13)  &  25.563(87) &  0.89720(97)   \\ \hline
14 &  0.32821(88) &  0.66635101(89) &  34.73(12)  &  0.89827(83)   \\ \hline
16 &  0.3347(13)  &  0.66644968(86) &  45.64(24)  &  0.89901(71)   \\ \hline
18 &  0.3443(13)  &  0.66651002(61) &  56.71(29)  &  0.89938(66)   \\ \hline
20 &  0.3494(17)  &  0.66654987(56) &  70.10(42)  &  0.89985(62)   \\ \hline
\end{tabular}
\caption{Values for $L_t=2$.}
\end{table}

\begin{table}[h!]
\begin{tabular}{|l|l|l|l|l|}
\hline 
$L_s$ &  max $C_V$ & min $B_4$ & max $\chi$ & $\beta_{pc}$\\ \hline
18 &  4.733(29)  &  0.6662128(28) &  139.59(89) & 1.009329(48)  \\ \hline
24 &  7.370(52)  &  0.6663686(22) &  270.4(1.8) & 1.009474(23)  \\ \hline
30 &  10.606(82) &  0.6664471(17) &  461.7(3.1) & 1.009542(14)  \\ \hline
36 &  14.82(11)  &  0.6664891(14) &  733.3(5.2) & 1.0095727(93) \\ \hline
42 &  19.87(22)  &  0.6665167(17) &  1089(11)   & 1.0095877(76) \\ \hline
48 &  26.02(27)  &  0.6665350(13) &  1560(15)   & 1.0095976(49) \\ \hline
54 &  32.92(46)  &  0.6665498(16) &  2119(27)   & 1.0096083(45) \\ \hline
\end{tabular}
\caption{Values for $L_t=6$.}
\end{table}

\begin{table}[h!]
\begin{tabular}{|l|l|l|l|l|}
\hline 
$L_s$ &  max $C_V$ & min $B_4$ & max $\chi$ & $\beta_{pc}$\\ \hline
30 &  19.98(11) &  0.6663168(19)  &  566.1(2.8)  & 1.0102814(72)   \\ \hline
33 &  25.45(12) &  0.6663319(15)  &  755.8(3.3)  & 1.0102910(49)   \\ \hline
36 &  31.98(12) &  0.6663395(11)  &  998.3(3.3)  & 1.0103034(34)   \\ \hline
39 &  40.40(12) &  0.6663449(10)  &  1297.0(3.6) & 1.0103134(25)   \\ \hline
42 &  50.00(14) &  0.66634863(97) &  1655.2(4.5) & 1.0103190(21)   \\ \hline
\end{tabular}
\caption{Values for $L_t=7$.}
\end{table}

\begin{table}[h!]
\begin{tabular}{|l|l|l|l|l|}
\hline 
$L_s$ &  max $C_V$ & min $B_4$ & max $\chi$ & $\beta_{pc}$\\ \hline
18 &  8.989(68)  &  0.6660313(45) &  125.86(74) &  1.010453(24)   \\ \hline
21 &  12.28(10)  &  0.6661217(46) &  188.0(1.3) &  1.010528(19)   \\ \hline
24 &  16.599(92) &  0.6661728(27) &  274.8(1.2) &  1.0105830(97)   \\ \hline
27 &  22.70(12)  &  0.6661929(25) &  397.1(1.9) &  1.0106238(71)   \\ \hline
30 &  30.12(15)  &  0.6662085(23) &  553.4(2.4) &  1.0106372(48)   \\ \hline
33 &  39.50(19)  &  0.6662155(21) &  750.2(3.1) &  1.0106464(39)   \\ \hline
\end{tabular}
\caption{Values for $L_t=8$.}
\end{table}

\end{document}